\documentclass[aip,jmp,amsmath,amssymb,reprint,superscriptaddress,nobibnotes]{revtex4-1}
\usepackage{graphicx}

\begin{document}

\title{Casimir probe based upon metallized high Q SiN nanomembrane resonator}

\author{Daniel Garcia-Sanchez}
\affiliation{Department of Electrical Engineering, Yale University, New Haven, CT 06520}

\author{King Yan Fong}
\affiliation{Department of Electrical Engineering, Yale University, New Haven, CT 06520}

\author{Harish Bhaskaran}
\affiliation{Department of Electrical Engineering, Yale University, New Haven, CT 06520}

\author{Steve Lamoreaux}
\affiliation{Department of Physics, Yale University, New Haven, CT 06520}

\author{Hong X. Tang}
\affiliation{Department of Electrical Engineering, Yale University, New Haven, CT 06520}

\date{\today}

\begin{abstract}
We present the instrumentation and measurement scheme of a new Casimir force probe that bridges
Casimir force measurements at microscale and macroscale. A metallized high Q
silicon nitride nanomembrane resonator is employed as a sensitive force probe. The
high tensile stress present in the nanomembrane not only enhances the quality factor
but also maintains high flatness over large area serving as the bottom
electrode in a sphere-plane configuration. A fiber interferometer is used to
readout the oscillation of the nanomembrane and a phase-locked loop scheme is
applied to track the change of the resonance frequency. Because of the high
quality factor of the nanomembrane and the high stability of the setup, a
frequency resolution down to $2\times10^{-9}$ and a corresponding force gradient
resolution of 3 $\mu$N/m is achieved. Besides sensitive measurement of Casimir
force, our measurement technique simultaneously offers Kelvin probe measurement
capability that allows \emph{in situ} imaging of the surface potentials.

\end{abstract}


\maketitle

\section{Introduction}
The first experiment on Casimir force measurement was conducted in 1958 by
Sparnnay\cite{MJSparnaayPhysicA1958} with macroscopic metal plates. In this
experiment the alignment of two parallel plates posed great challenges. In 1997,
the accuracy of the measurement was significantly improved by using a torsion
pendulum balance\cite{SKLamoreauxPRL1997}. The alignment was simplified by
replacing one of the plates by a sphere characterized by its radius of curvature
$R$. The distance was determined by measuring the capacitance between the two
plates.
In 2011 Sushkov~\emph{et al.} used an improved version\cite{AOSushkovNatPhys2011} of the torsion pendulum balance
to demonstrate that the permitivity is well described by the Drude model in the range
$0.7$ to $7 \mathrm{\mu m}$.

Moving towards smaller device sizes, in 1998 Mohindeen \emph{et al.} measured the
Casimir force also in a sphere-plane geometry in the range of $0.1$ to
$0.9\,\mu\textrm{m}$, but with an Atomic Force
Microscope~(AFM) as a sensitive force
transducer\cite{UMohideenPRL1998}. An aluminium coated polystyrene sphere
was attached to an AFM cantilever. The force was measured between the sphere and
an aluminium coated sapphire plate by measuring the deflection of the AFM
cantilever. Using a similar AFM approach repulsive Casimir force was
demonstrated in a fluid medium by Munday~\emph{et al.}\cite{JNMundayNature2009}.

Microelectromechanical systems (MEMS) have also been exploited to study the Casimir
force. In one of the first experiments with MEMS it was shown that a cantilever
could irreversibly stick to an electrode due to the Casimir
force\cite{EBuksEPL2001}. A microelectromechanical torsion
oscillator\cite{HBChanScience2011} was used by Chan~\emph{et al.} to measure
with high accuracy the Casimir force in the range of $0.1$ to $1\,\mathrm{\mu m}$
and later to demonstrate non-linear oscillations in the micromechanical torsion
oscillator due to the Casimir force\cite{HBChanPRL2001}. In 2002 Bressi~\emph{et
al.} measured the Casimir force between parallel plates\cite{GBressiPRL2002} in
the range $0.5$ to $3\,\mathrm{\mu m}$.  In this experiment the Casimir force was
measured in an area of $1.2\,\mathrm{mm} \times 1.2\,\mathrm{mm}$.

Here we describe the development of a new type of Casimir force probe that
bridges the Casimir force measurement at macroscopic scale and microscopic
scale. Results of the measurement of this force are presented
elsewhere\cite{DGarciaPRL2012}. In a sphere-plane geometry, the planar plate, which
serves as force transducer, is a high Q SiN nanomembrane resonator. The sphere
is a gold coated glass sphere of millimeter size. This approach combines the
benefits of free space manipulability, large surface area and high force
sensitivity of nanomembranes. The high force sensitivity allows precision
measurement of the shift of the resonance frequency of a nanomembrane resonator.
The shift is proportional to the gradient of the forces across the gap which
could have newtonian origin (such as electrostatic force) or non-newtonian
origin (such as the Casimir force). These dynamic techniques\cite{HBChanPRL2001,WPerniceOctExpr2010} have several
advantages over the static force measurement techniques which are very
susceptible to environmental instabilities such as the seismic vibrations and
the $1/f$ noise that is inherent in the measurement system.

\begin{figure*}[ht]
\begin{center}
\includegraphics{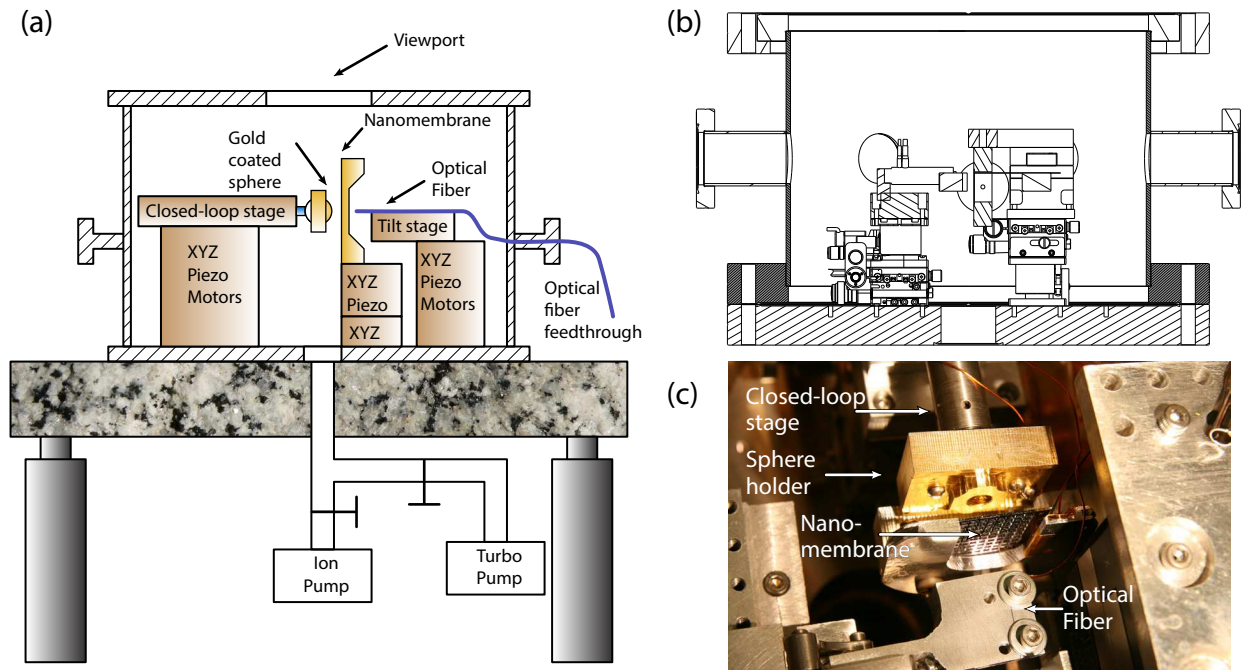}
\end{center}
\caption{\textbf{(a)} Schematic of the isolation system, the vacuum chamber and the positioning system inside the vacuum chamber.
\textbf{(b)} Cut view of the chamber and the probe.
\textbf{(c)} Isometric image of the sample mounting subsystem and fiber mounting subsystem. }\label{fig:VacuumSystem}
\end{figure*}

\section{Instrumentation}

\subsection{Vacuum system}
Figures \ref{fig:VacuumSystem}(a) and \ref{fig:VacuumSystem}(b) show a schematic illustration of the Casimir
probe setup which is installed in a high vacuum chamber of $75\,\mathrm{kg}$. To
minimize the environmental impact, the chamber is placed on top of a granite
table ($1500\,\mathrm{kg}$) with pneumatic vibration isolation. Further, a wood triangle
of $2.5\,\mathrm{in}$ in thickness is inserted between the vacuum chamber and
the granite table to achieve optimum damping of environmental vibrations. A
turbo pump system is used to pre-pump the vacuum chamber. When the chamber
pressure is below $10^{-6}\,\textrm{Torr}$, an ion pump is turned on and the
turbo pump is switched off and detached from the vacuum system in order to
remove the pump induced vibrations. The fiber feedthrough is constructed by
modifying a Swagelok connector fitted with a teflon
ferrule\cite{RIEricApOpt1998}. All electrical feedthroughs are made ultra-high vacuum (UHV)
compatible by Accu-Glass Products, Inc. We finally leak-test the vacuum system
with a helium leak detector, and the leaking rate is
below~$10^{-9}\,\textrm{Torr}$.

The positioning system inside the vacuum chamber consists of 13 motorized
stages. The sphere, the nanomembrane, and the fiber are separately mounted on
three specifically designed holders, each of which is positioned by an XYZ stage
controlled by picomotors\cite{refpicomotoruhv} (See Fig.~\ref{fig:VacuumSystem}(a)).
An additional 3-axis piezoelectrically actuated stage\cite{refnanocube} is
mounted on the nanomembrane stage to achieve sample scanning and to fine adjust
fiber-to-membrane distances. The alignment of the fiber against nanomembrane is
adjusted by a manual tip-tilt stage. The sphere is brought to approach the
nanomenbrane by a preloaded closed-loop piezo actuator with sub-nanometer
resolution ($0.3\,\textrm{nm}$)\cite{RefPreloadedPiezo}. All the stages are made vacuum compatible and
carefully cleaned before installation. The movement of the stages can be
carefully monitored using a telescope through the viewport at the top of the
vacuum chamber. During the measurements the viewport is fully covered with metal
foil to avoid ambient light which is found to introduce noise in the
measurements.

\subsection{Preparation of the spherical plate}
\begin{figure}[ht]
\begin{center}
\includegraphics{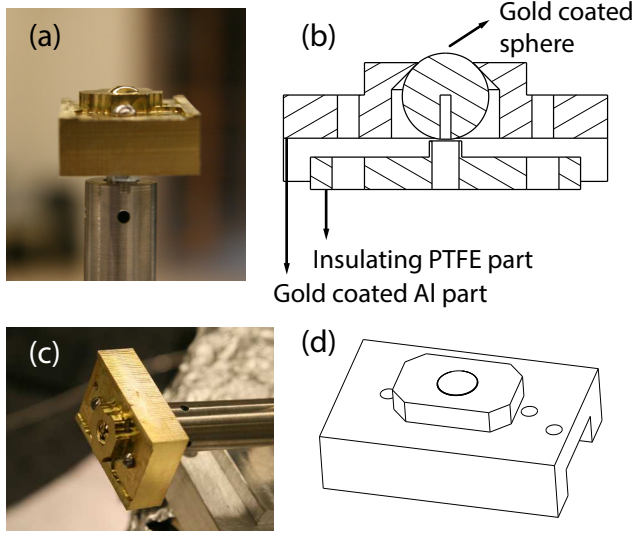}
\end{center}
\caption{Casimir sphere and its holder \textbf{(a)} Side-view photograph.
\textbf{(b)} Cross-sectional view of the drawing assembly.
\textbf{(c)} Perspective view. \textbf{(d)} Isometric view of the drawing.}\label{fig:HolderBall}
\end{figure}

Figure~\ref{fig:HolderBall} shows the spherical plate and its holder. The sphere
is prepared by coating a fused silica with $200\,\textrm{nm}$ of gold using
$10\,\textrm{nm}$ of titanium as adhesion layer using an electron beam physical
vapor deposition.  The sphere\cite{RefSpherePRL} has a radius $R =
4\,\mathrm{mm}\pm\,2.5\,\mathrm{\mu m}$. The holder of the
sphere is machined in aluminium and sputtered with gold to reduce the contact
potential between the sphere and the holder. To ensure good electrical contact
between the sphere and the holder a polytetrafluoroethylene (PTFE) stub is used
to press the sphere against the holder, as shown in Fig.~\ref{fig:HolderBall}(b). The
PTFE piece is also used to electrically isolate the
sphere from the closed-loop piezo actuator and the vacuum chamber. Static charge
might accumulate on the PTFE surface. But this is not a concern since the PTFE
piece is fully electrically shielded from the measurement surface.

\subsection{Nanomembrane fabrication}
In the sphere-plane configuration, the planar plate used here is a
nanomembrane fabricated from tensile-stressed silicon nitride which also serves
as the force transducer. Membrane resonators made of tensile-stressed silicon
nitride are known to exhibit very high mechanical Q (in the order of $10^6$ at
room temperature)\cite{JDThompsonNature2008,IWilsonRaePRL2011}. The Q can be
as high as $~10^5$ even after metallization\cite{PLYuPRL2012}, which is expected
to introduce extra material loss and clamping loss. From a transducer point of
view, high Q implies low thermomechanical noise and so high force sensitivity.
Here, we make use of this advantageous property of the high Q nanomembrane to
achieve precision measurement of Casimir force.

\begin{figure}[t]
\begin{center}
\includegraphics{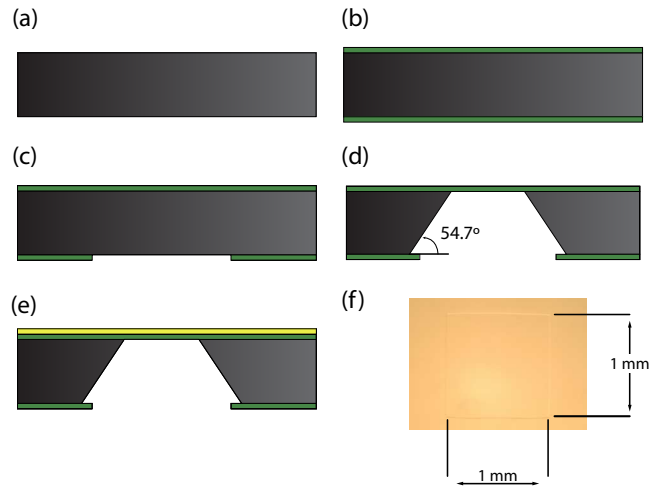}
\end{center}
\caption{Fabrication process of the nanomembrane. \textbf{(a)} Bare Si wafer.
\textbf{(b)} Deposition of 330nm LPCVD silicon nitride. \textbf{(c)} CHF$_{3}$/O$_{2}$ RIE of silicon nitride \textbf{(d)} Anisotropic KOH etching of silicon. \textbf{(e)} Ebeam evaporation of Au/Ti layer. \textbf{(f)} Optical image of the nanomembrane.}\label{fig:FabricationMembrane}
\end{figure}

Fabrication process of the nanomembrane is described as follows. We start with a
$400\,\mathrm{\mu m}$ thick, (100)-oriented, double side polished silicon wafer
(doped p-type with resistivity quoted to be 10 - 20 $\Omega$cm). A layer of 330
nm silicon nitride is grown on both sides of the wafer by low pressure chemical
vapor deposition (LPCVD). The silicon nitride Si$_x$N$_y$ used here is not
stoichiometric but has a silicon-to-nitrogen ratio $x:y$ larger than $3:4$. When
deposited on a silicon substrate, the film has a built-in tensile stress of 450
MPa, which is relatively lower than that of a stoichiomeric silicon nitride. A
layer of photoresist (Shipley S1813) was spun on both sides of the wafer. The
photoresist on the front side serves as a protective layer. Square openings were
defined in the photoresist on the backside of the wafer using photolithography
and the silicon nitride layer was etched by reactive ion etching (RIE) with a
mixture of CHF$_3$ and O$_2$ gases. After stripping off the photoresist, an
anisotropic wet etch of potassium hydroxide (KOH) solution (30\%, 85 $^\circ$C
for 7 hours) was used to etch through the Si substrate. During the KOH etching
process a ProTEK-B3 (Brewer Science) coating was used to protect the silicon
nitride layer on the top side of the wafer in order to ensure the high quality
of the membrane and increase the yield. After the KOH etch, the coating was
removed in a bath of N-Methyl-2-pyrrolidone (NMP) at 80 $^\circ$C and a
subsequent O$_2$ plasma ashing. Finally $10\,\mathrm{nm}$ of titanium and $200\,\mathrm{nm}$ of gold
were evaporated on top of the membrane, at a rate of 1 \AA/s. Fig.~\ref{fig:FabricationMembrane}(f)
shows the typical optical image of a gold coated
nanomembrane. Due to the net tensile stress in the film, the sample remains flat
across the whole nanomembrane surface with RMS roughness of $3\,\mathrm{nm}$ (measured by AFM).
Figure \ref{fig:FabricationMembrane} summarizes the nanomembrane fabrication
process.

\subsection{Fiber interferometer and its calibration}
\begin{figure}[ht]
\begin{center}
\includegraphics{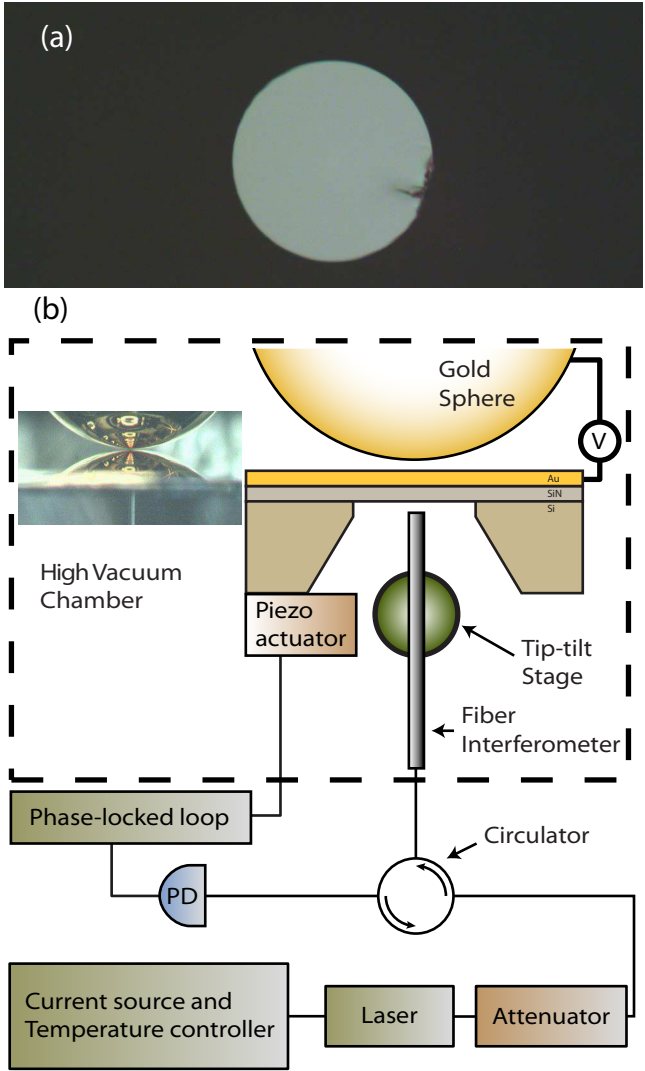}
\end{center}
\caption{\textbf{(a)} Surface of the cleaved fiber.
\textbf{(b)} Schematic diagram of the fiber-optic interferometer and the detection and actuation scheme.}\label{fig:schemafiberinterf}
\end{figure}

We first describe the fiber interferometer employed to measure the resonance of
the nanomembrane. The fiber is cleaved after it has been introduced into the
chamber to produce a clean cleaved surface as shown in
Fig.~\ref{fig:schemafiberinterf}(a). Figure~\ref{fig:schemafiberinterf}(b) displays
the schematic
of the fiber measurement scheme. A diode laser\cite{RefLaser} ($\lambda =
1310\,\mathrm{nm}$) with very low frequency noise is used as a light source. The output
power of the laser is stabilized with a current and temperature
controller\cite{lasercontroller} and is set to $9.7\,\mathrm{mW}$ (operating
current $40\,\mathrm{mA}$) which corresponds to the setpoint with the lowest
frequency noise. Then the light is attenuated to $500\,\mu W$ and sent to the
nanomembrane through a circulator\cite{RefCirculator}. The light reflected from the nanomembrane returns to
the circulator and is collected with a photodetector.

\begin{figure}[ht]
\begin{center}
\includegraphics{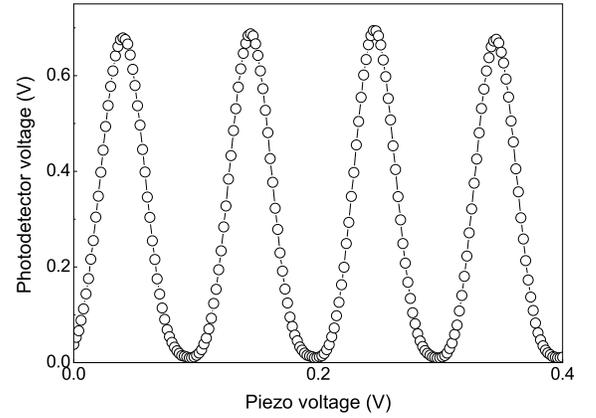}
\end{center}
\caption{Interference pattern.}\label{fig:interfpattern}
\end{figure}

When the distance between the fiber and the nanomembrane is changed an
interference pattern is obtained (see Fig.~\ref{fig:interfpattern}). In order to
improve the extinction, the fiber axis is fine adjusted by a tip-tilt stage~(see
Fig.~\ref{fig:schemafiberinterf}(b)). The maximum extinction is obtained when the
fiber is perpendicular to the surface of the nanomembrane. We record the
photodetector signal as the nanomembrane position is varied. The maximum
sensitivity is obtained where the slope is maximal in the interference
pattern~(see Fig.~\ref{fig:interfpattern}), at which point a small change in
distance $\Delta d$ results in a change of photodetector signal $\Delta V$,
which is related to a change by
\begin{equation}
 \Delta d = \frac{\lambda}{V_{p-p}4\pi}\Delta V
\end{equation}
where $V_{p-p}$ is the peak-to-peak voltage of the interference pattern and $\lambda$ the wavelength of the laser.

\subsection{Coarse approach betwen the fiber and the nanomembrane}
When the fiber is brought to approach the nanomembrane, the distance between the
fiber and the nanomembrane is measured in real time to avoid physical contact
and breaking of the nanomembrane. We achieve this at each distance by sweeping
the wavelength using a tunable laser\cite{refsweeplaser}, and recording the maxima
and minima of the interferometer signal.  The distance is given by
\begin{equation}
d = \frac{N}{2}\left(\frac{1}{\lambda_i}-\frac{1}{\lambda_f}\right)^{-1}
\end{equation}
where $d$ is the distance between the fiber and the nanomembrane, $N$ the number
of peaks in the interference pattern in the wavelength regime, $\lambda_i$ the
starting wavelength and $\lambda_f$ the final wavelength.

In order to collect the maximum amount of light reflected by the nanomembrane
the fiber has to be very close to the nanomembrane because of the relative small
numerical aperture of the single mode fiber. With fiber-membrane distance
monitored in real time, the fiber is brought to the closest yet safe distance
from the nanomembrane by operating the picomotors. During the approach the fiber
is aligned several times with a tip-tilt stage\cite{RefRotationalStageRSI}~(see
Fig.~\ref{fig:schemafiberinterf}(b)). When the fiber is aligned
the extinction of the interference pattern
is maximized. By doing this, we improve the sensitivity of the interferometer
and are able to approach to a very close distance. We then move the nanocube Z
stage to fine tune the fiber-membrane distance so that maximum sensitivity is
achieved in our fiber interferometer.

\subsection{Setup stability evaluation}
\begin{figure}[ht]
\begin{center}
\includegraphics{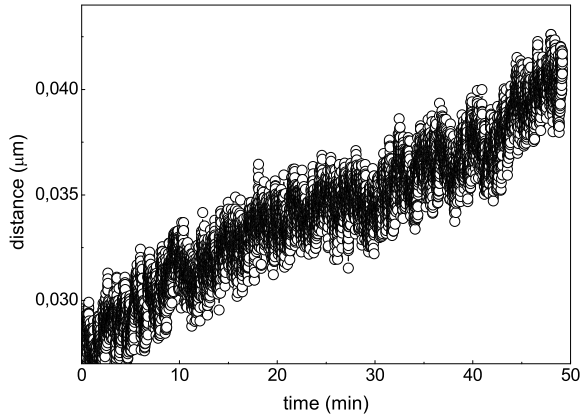}
\end{center}
\caption{Interferometer output for 45 min record.}\label{fig:timeinterferometer}
\end{figure}

In order to minimize the mechanical drift of the fiber interferometer and the
apparatus the room temperature is kept at $20.0\pm 0.1^\circ\textrm{C}$. The big
mass of the vacuum system further filters out the variations of the temperature
of the room. We characterize the stability of the setup by monitoring the
interferometer signal overnight. As shown in Fig.~\ref{fig:timeinterferometer}
the measured drift of the fiber-to-membrane distance is about
$0.26\,\textrm{nm}/\textrm{min}$ which is probably due to the thermal expansion
of the apparatus and stages. As explained in the next section one single
measurement (a parabola fitting) takes about $800\,\mathrm{ms}$ which corresponds to a
drift of $3\,\mathrm{pm}$ which is negligible when compared with
other sources of error, such as the  determination of the
distance and the measurement of the resonance frequency of the nanomembrane.

\subsection{Nanomembrane characterization}
\begin{figure}[ht]
\begin{center}
\includegraphics{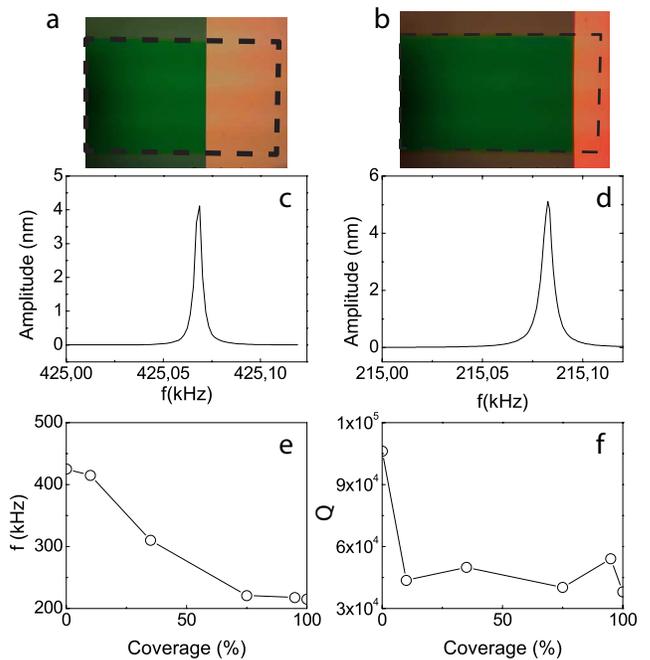}
\end{center}
\caption{\textbf{(a)} Optical image of a nanomembrane with a 36\% of the surface coverered by gold.
\textbf{(b)} Optical image of a nanomembrane with a 12\% of the surface coverered by gold.
\textbf{(c)} Frequency response of a nanomembrane not covered by gold.
\textbf{(d)} Frequency response of a nanomembrane fully covered by a layer of gold.
\textbf{(e)} Resonance frequency of the nanomembrane vs the amount of the surface covered by a layer of gold.
\textbf{(f)} Quality factor of the nanomembrane vs the percentage of gold coverage.}\label{fig:coverage}
\end{figure}

We measured the frequency response of the SiN gold coated nanomembranes and
studied how the quality factor is modified when they are covered with gold. The
nanomembrane is excited with an oscillating voltage $V_{AC}$ which is applied to
the piezo actuator and the response is measured with the fiber interferometer.
The amplitude-frequency response of the nano-membrane was obtained by measuring
the oscillation amplitude of the nano-membrane and sweeping the driving
frequency excitation. The frequency response was measured in a set of samples
where the amount of surface covered by gold changes~(see
Figs.~\ref{fig:coverage}(a) and \ref{fig:coverage}(b)). It goes from fully covered
by a layer of $200\,\mathrm{nm}$ of gold to not covered at all.
Fig.~\ref{fig:coverage}(c) shows the response curve of a nanomembrane which is not
covered by gold and Fig.~\ref{fig:coverage}(d) shows the response curve of a
nanomembrane which is fully covered by a layer of $200\,\mathrm{nm}$ of gold.
As we increase the amount of surface coverage the resonance frequency decreases
because the total mass of the nanomembrane increases, this is shown in
Fig.~\ref{fig:coverage}(e). When the nanomembrane is covered with some gold the
quality factor is reduced, but we have not observed a deterministic
trend of the quality factor with regards to the gold coverage as shown in
Fig.~\ref{fig:coverage}(f). Hence we choose fully covered membrane in our
following measurements.

\subsection{Frequency drift and resolution}
\begin{figure}[t]
\begin{center}
\includegraphics{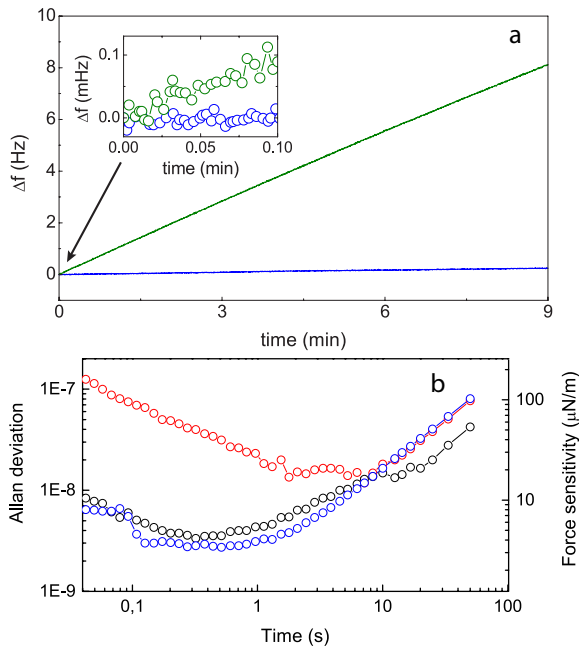}
\end{center}
\caption{\textbf{(a)} Frequency drift with blind viewports~(blue) and with transparent viewports~(green). The inset shows a zoom-in of the data.
\textbf{(b)} Left axis shows the Allan variance vs the integration time for several drive voltages: $500\,\mathrm{\mu V}$ (red), $10\,\mathrm{mV}$
(black), and $50\,\mathrm{mV}$ (blue). The right axis shows the equivalent gradient force sensitivity.}\label{fig:temproomdriftfreq}
\end{figure}

The frequency of the nanomembrane is tracked using a phase-locked loop (PLL)
scheme. The output of the photodiode is compared with a reference signal using a
lock-in amplifier that has a built-in PLL capability\cite{reflockin}. We use the
frequency shift of the nanomembrane to calculate the Casimir and the
electrostatic forces~(see section~\ref{sec:principle}). The temperature drift of
the nano-membrane can also shift the resonance frequency. As
mentioned before the temperature of the room is stabilized to $20.0\pm
0.1^\circ\textrm{C}$ and also the big mass of the chamber stabilizes the
temperature of the nanomembrane. The ambient light can also substantially heat
up the nanomembrane and shift the resonance frequency, therefore all the
viewports of the chamber are blinded. When ambient light is allowed to enter
into the chamber the frequency drift is about $1\,\mathrm{Hz}/\mathrm{min}$, and
when the viewports are blinded
the frequency drift is $0.26\,\mathrm{mHz}/\mathrm{min}$ (see
Fig.~\ref{fig:temproomdriftfreq}(a)). The frequency drift is constant during a
period of time much longer than a typical measurement, which is shorter than one
second. Before any measurement the frequency drift is calibrated and the
measurements are corrected according to this calibration.
By optimizing all the relevant conditions, we achieve a frequency resolution of $\Delta f/f_0=2\cdot10^{-9}$ which allows to measure
a force gradient of $3\,\mathrm{\mu N/m}$ (see Fig.~\ref{fig:temproomdriftfreq}(b)).
As shown in the Allan deviation the drift becomes relevant above a measurement time of $1\,\mathrm{s}$ which is smaller than the measurement time of one parabola.

\section{Measurement scheme}

\subsection{Principle}\label{sec:principle}
To measure the Casimir force we use the resonance frequency of the nanomembrane as
a transducer which is very sensitive to the force gradient. When the
nanomembrane approaches the sphere, it experiences a force which has
contributions from the Casimir Force as well as the electrostatic force. This
force is distributed over the whole nanomembrane and the force per unit of
surface can be expressed as $\mathbf{F_e}= F_{e}\,\mathbf{F_u}$, where $F_{e}$
represents the magnitude of the force and $\mathbf{F_u}$ the normalized spatial
distribution of the force. Now if we consider the first 3 terms of the Taylor
expansion of the external force $F(z)$ about the distance $z_0$, the equation of
motion of the nanomembrane becomes

\begin{eqnarray}
 \ddot{z}+2\gamma\dot{z}+\omega^2_m (z-z_0)&=&\frac{\omega_m^2\langle \mathbf{F_u},\mathbf{u_d}\rangle}{k_{\mathrm{m}}}F_d(z_0)e^{i \omega_d t} \nonumber\\
&&+\frac{\omega_m^2\langle \mathbf{F_u},\mathbf{u_d}\rangle}{k_{\mathrm{m}}}F'(z_0)(z-z_0) \nonumber\\
&&+\frac{\omega_m^2\langle \mathbf{F_u},\mathbf{u_d}\rangle}{k_{\mathrm{m}}}\frac{F''(z)}{2} (z-z_0)^2\nonumber\\
&&+\frac{\omega_m^2\langle \mathbf{F_u},\mathbf{u_d}\rangle}{k_{\mathrm{m}}}\frac{F'''(z)}{6} (z-z_0)^3\nonumber\\
\label{eq:equationmv}
\end{eqnarray}
where $\omega_m=2\pi f_m$ is the fundamental angular frequency of the
nanomembrane, $\omega_d$ the angular driving frequency, $\gamma$ is the damping
coefficient, $k_{m}$ the stiffness of the nanomembrane and $\mathbf{u_d}$ the
eigenvector of the first eigenmode. The product $\langle
\mathbf{F_u},\mathbf{u_d}\rangle$ accounts for the overlap between the force and
the eigenmode. We can then define the effective stiffness as:
\begin{equation}
 k_\mathrm{eff}=\frac{k_{m}}{\langle \mathbf{F_u},\mathbf{u_d}\rangle}\label{eq:effstiffness}
\end{equation}
In the case of the first eigenmode the overlap between the force and the
eigenmode will be maximum when the sphere is centered on top of the
nanomembrane, and the stiffness will be minimum from the point of view of the
external force.

The amplitude of motion of the resonator is set to be in the linear regime but
there is some contribution from high order derivatives of the external force. If
we consider a small non-linearity the resonance
frequency can be approximated by
\begin{eqnarray}
 \Delta f&=&-\frac{f_m}{2 k_{eff}}\left(F'(z_0)+\frac{A^2_{\mathrm{rms}}}{4}F'''(z_0)\right)\nonumber\\
 &=&-\frac{f_m}{2 k_{eff}}F'_{\mathrm{a}}(z_0)\label{eq:freqshiftforce}
\end{eqnarray}
where $F'_{\mathrm{a}}$ is the apparent force and $A_{\mathrm{rms}}$ the RMS
amplitude of motion of the resonator. The contribution of the third derivative
of the force to the frequency shift can be considered as a correction and is
equivalent to the corrections for the roughness and the fluctuations in
plate separation that have been considered in previous experiments\cite{SKLamoreauxPRA2010,AOSushkovNatPhys2011}.

\subsection{Measurement procedure}
When a difference of voltage $V$ is applied between the sphere and the
nanomembrane an electrostatic force appears and is given by
\begin{equation}
 F_{e}=-\pi\epsilon_0 R \frac{(V-V_m)^2}{z-z_0}
\end{equation}
where $\epsilon_0$ is the vacuum permittivity, $R$ the radius of the sphere,
$z-z_0$ the absolute distance between the sphere and the nanomembrane, $z$ the
relative distance which is set with the closed-loop piezo actuator and $V_m$ the
residual contact potential.

\begin{figure}[t]
\begin{center}
\includegraphics{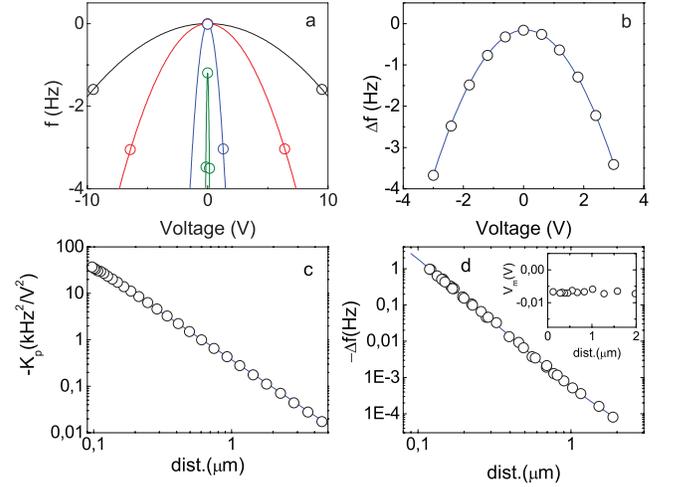}
\end{center}
\caption{\textbf{(a)} Parabolas at different target distances: $9\,\mu\textrm{m}$ (blue), $4\,\mu\textrm{m}$ (red), $1\,\mu\textrm{m}$ (green), $120\,\textrm{nm}$ (black).
\textbf{(b)} Frequency vs Voltage at a fix distance. The data can be nicely fitted to a parabola.
In this measurement $\chi^2=1.05$.
\textbf{(c)} Extracted $K_p$ (curvature of the parabola).
\textbf{(d)} Extrated $\Delta f_0$ which has contributions from the Casimir force and the residual electrostatic force\cite{DGarciaPRL2012}. Inset, the contact potential $V_\textrm{m}$ vs
the distance with the sphere.}\label{fig:parabolas}
\end{figure}

As shown in Figs.~\ref{fig:parabolas}(a) and \ref{fig:parabolas}(b), the
contribution of the electrostatic force to the frequency shift can be fitted to
a parabola characterized by a curvature $K_p$. Therefore at each distance
between the sphere and the nanomembrane it can be found that the frequency shift
is given by\cite{MBrownHayesPRA2005}
\begin{eqnarray}
 f^2 &=& f_0(z)^2+K_p(z)(V-V_m(z))^2\nonumber\\
&=& f_0(z)^2-\frac{\epsilon_0\pi R f^2_m}{k_\mathrm{eff}(z-z_0)^2}(V-V_m(z))^2\label{eq:freqshiftfquad}
\end{eqnarray}
In this equation $K_p(z)(V-V_m(z))^2$ accounts for the electrostatic force that
can be nulled by applying a voltage between the sphere and the nanomembrane.
Note that $V_\mathrm{m}$ is not necessarily constant with the relative position between the membrane
and the sphere and it has been shown that such a variation of $V_\mathrm{m}$ can cause
an additional electrostatic force\cite{DGarciaPRL2012,WJKimPRA2010}.
$f_0^2$ accounts for the Casimir force and the electrostatic force that can not
be cancelled. To calculate the magnitudes $K_p(z)$, $f_0(z)$ and $V_m(z)$ the
distance $z-z0$ between the sphere and the nanomembrane is fixed and three
different voltages $V_1$, $V_2$ and $V_3$ are applied between the sphere and the
nanomembrane. For each voltage a frequency shift is measured: $f_1$, $f_2$ and
$f_3$. With these measurements we can calculate the desired magnitudes and their
dependence with the distance $z$.
\begin{eqnarray}
     V_m&=& \frac{1}{2}\frac{f_3^2(V_2^2 -V_1^2 ) + f_2^2(V_1^2 - V_3^2) + f_1^2(V_3^2 -V_2^2)}{f_3^2(V_2-V_1) + f_2^2(V_1 - V_3) +   f_1^2(V_3-V_2)}\nonumber\\
    \\
    f_0&=&\sqrt{\frac{f_2^2(V_1-V_m)^2-f_1^2(V_2-V_m)^2}{(V_1-V_m)^2-(V_2-V_m)^2}}\\
    K_c&=&\frac{f_1^2-f_0^2}{(V_1-V_m)^2}
\end{eqnarray}

In order to compensate for the drift in frequency of the nanomembrane, the
measured quantities $f_1$, $f_2$ and $f_3$ have been corrected.
By fitting $K_p(z)$ to the expected model\cite{WJKimPRL2009} the quantities
$k_\mathrm{eff}$ and $z_\mathrm{off}$ can be obtained. With these quantities we
can have an absolute calibration of the distance.
Fig.~\ref{fig:parabolas}(c) shows a typical measurement of $K_p(z)$ and the fit with
$\chi^2$ of $1.04$. The fit gives as a result an effective mass of $7.5\,\mathrm{\mu g}$.
The expected theoretical effective mass for this particular nanomembrane is $1.2\,\mathrm{\mu g}$.
In the measurement of Fig.~\ref{fig:parabolas}(c) the nanomembrane appears to be heavier because
in this measurement the sphere is not centered on top of the nanomembrane.

$V_m$ is voltage that has to be applied to minimize the electrostatic
contribution to the force. This voltage may not be constant with the distance
between the sphere and the nanomembrane\cite{WJKimPRA2010,SLamoreauxArXiv2008}.
The quantity $f_0$ has contributions from the Casimir force but also from the
residual electrostatic force that can not be completely canceled even by setting
the external voltage $V$ to $V_m$. This force has the origin in the interplay
between the surface patches and the curvature of the sphere.
\begin{equation}
 \Delta f_0 = - \frac{f_m}{2 k_{eff}}\left(\frac{dF_c(z)}{dz}+\frac{dF^{el}_{res}(z)}{dz}\right)\label{eq:Deltaf0}
\end{equation}
where $F_c$ is the Casimir force and $F^{el}_{res}(z)$ is the residual
electrostatic force\cite{WJKimPRA2010,DGarciaPRL2012}.
To reduce the error introduced by the drift of the frequency in  $\Delta f_0$,
two parabola measurements are done, one at a very far distance and another one
at the target distance $d_0$. This gives as a result $f_{0\mathrm{f}}$ and
$f_{0\mathrm{c}}$ respectively. $\Delta f_0$ is equal to the subtraction between
these two frequencies.  This procedure is repeated several times to improve the
signal to noise ratio of $\Delta f_0$ at $d_0$. The value $\Delta f_0$  is the
average of all the $\Delta f_0$ measured at that particular distance.
Fig.~\ref{fig:parabolas}(d) shows a typical measurement of $\Delta f_0(z)$, and
the inset shows the contact potential $V_\textrm{m}$ vs the distance with the sphere.
Note that Fig.~\ref{fig:parabolas}c and Fig.~\ref{fig:parabolas}d are based on different sets of data.
The calibration curve of Fig.~\ref{fig:parabolas}c is obtained using the displacement data of the closed loop piezo actuator.
To be insensitive to the drift of the system this data has been taken quite fast.
Once the system is calibrated we can measure $\Delta f$ and $V_m$ by doing repetitions of the same measurement in order to improve the accuracy (see Fig.~\ref{fig:parabolas}d). In this case the displacement data of the piezo actuator is discarded and the distance is calculated using the measured $K_p$. This method is very accurate and insensitive to the mechanical drift.

\subsection{Surface potential measurements}
\begin{figure}[ht]
\begin{center}
\includegraphics{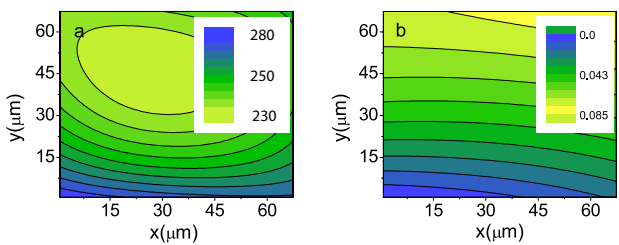}
\end{center}
\caption{\textbf{(a)} Measured stiffness of a nanomembrane in $\mathrm{N}/\mathrm{m}$.
The stiffness at the center is minimum and correspond for this particular sample to $230\,\mathrm{N/m}$.
\textbf{(b)} Measured surface potential in Volts.
In this sample there is a strong variation of the surface potential across the surface.}\label{fig:keff}
\end{figure}

Scanning Kelvin probe principle is a very powerful noninvasive
technique\cite{NARobertsonClassQuanGrav2006,MNonnenmacherAPL1991,
AKikukawaAPL1995} that allows to measure the surface contact potential $\Delta
V$ between a conducting tip and a sample. The surface contact potential $\Delta
V $  is related to the difference of the work function between the sphere and
the nanomembrane
\begin{equation}
 \Delta V =-\frac{W_2-W_1}{e}
\end{equation}
where $e$ is the elementary charge and $W_1$ and $W_2$ are the work functions of
the tip and the sample respectively. The work function and as a consequence the
surface contact potential can vary along the surface due to oxides films or
adsorbed chemicals on the surface and produce the patch
potentials\cite{WJKimPRA2010,DGarciaPRL2012}. Our setup allows for in-situ
measurements of the surface potentials on the nanomembrane. The sphere can be
moved parallel to the nanomembrane using the XYZ Piezo System in order to image
the surface potential at a fix distance between the sphere and the
nano-membrane. The movement of the axes $X$ and $Y$ from the nanocube is not
completely parallel to the sphere, therefore when we move the nanomembrane using
the axis $X$ and $Y$ the distance between the sphere and the nanomembrane
changes. In order to measure the contact potential at a fix distance $z$, a
calibration curve is taken at each point $(x,y)$. Then using the calibration
curve the sphere is placed at a distance $z$ from the nanomembrane and the
contact potential is measured. Fig.~\ref{fig:keff}(b) shows the measurement of the
contact potential of a sample where the variations of the contact potentials are
large.
These contact potentials play an an important role in the precise measurement of
the Casimir force\cite{DGarciaPRL2012}. The spatial variations of the contact
potential create an electric field between the sphere and the nanomembrane which
originates a residual electrostatic force that cannot be completely canceled by
applying a a voltage between the sphere and the nanomembrane.

The effective stiffness $k_{\mathrm{eff}}$ also depends on the relative position
between the sphere and the nano-membrane. More specifically it depends on the
overlap between the mechanical mode of the nano-membrane and the
distribution of the electrostatic force~(see equation~\eqref{eq:effstiffness}).
The effective stiffness $k_{\mathrm{eff}}$ is minimum, when the overlap between
the mechanical eigenmode and the electrostatic force is maximum, that is when
the sphere is centered over the nanomembrane. When the sphere gets farther from
the center of the nanomembrane, the overlap between the eigenmode and the
electrostatic force becomes smaller and the effective stiffness betcomes bigger.
The spatial dependence of the effective stiffness can be used to center the
sphere over the top of the nanomembrane. Figure~\ref{fig:keff}(a) shows the
contour mapping of the measured stiffness at the center of a nano-membrane.

\section{Conclusion}
We have designed a Casimir force probe that allows \emph{in situ}
measurements of the contact potential and bridges the measurement of the Casimir
force at microscale and macroscale. The Casimir force measurement is based on a
sphere-plane geometry where the planar plate is a high Q silicon nitride
nanomembrane coated with gold. The probe can measure a force gradient down
to $3\,\mathrm{\mu N}/\mathrm{m}$.
We have also studied the dependence of the quality factor of the nanomembrane
with the amount of coated gold and we have observed that the SiN nanomembranes
coated with gold can retain a reasonably high quality factor. \emph{In situ}
scanned Kevin probe allows direct assessment of contact potentials and their
distributions, which are found to contribute significantly to the measured
position dependence forces in most samples we have tested.


\begin{thebibliography}{33}
\bibitem [{Ref({\natexlab{b}})}]{RefSpherePRL}%
  \BibitemOpen
  \href@noop {} {} \bibinfo {note} {The sphere is a fused
  silica ball lens from Edmund Optics with ref. NT67-388.}\BibitemShut {Stop}%
\end{thebibliography}

%

\end{document}